\documentclass[twocolumn,floatfix,amsmath,amssymb,showpacs]{revtex4}
\usepackage{graphicx}
\usepackage{dcolumn}
\usepackage{color}

\newcommand{\be}{\begin{equation}}
\newcommand{\ee}{\end{equation}}

\begin{document}
\title{Anomalous scaling of passive scalars in rotating flows}

\author{P. Rodriguez Imazio$^1$, P.D. Mininni$^{1,2}$}
\affiliation{$^1$ Departamento de F\'{\i}sica, Facultad de Ciencias Exactas y 
                  Naturales, Universidad de Buenos Aires and IFIBA, CONICET, 
                  Cuidad Universitaria, Buenos Aires 1428, Argentina \\
             $^2$ National Center for Atmospheric Research, P.O. Box 3000, 
                  Boulder, Colorado 80307, USA}
\date{\today}

\begin{abstract}
We present results of direct numerical simulations of passive scalar 
advection and diffusion in turbulent rotating flows. Scaling laws and the 
development of anisotropy are studied in spectral space, and in real space 
using an axisymmetric decomposition of velocity and passive scalar 
structure functions. The passive scalar is more anisotropic than the 
velocity field, and its power spectrum follows a spectral law consistent 
with $\sim k_\perp^{-3/2}$. This scaling is explained with phenomenological 
arguments that consider the effect of rotation. Intermittency is 
characterized using scaling exponents and probability density 
functions of velocity and passive scalar increments. In the presence of 
rotation, intermittency in the velocity field decreases more noticeably 
than in the passive scalar. The scaling exponents show good agreement 
with Kraichnan's prediction for passive scalar intermittency in 
two-dimensions, after correcting for the observed scaling of the second 
order exponent.
\end{abstract}

\maketitle

\section{Introduction}

A passive scalar is a quantity diluted in a fluid in such a low concentration 
that it does not affect the flow evolution, but that it is advected and 
diffused by the flow. Examples are given by colorant dye used in experiments, 
and aerosols and pollutants in small concentrations in the atmosphere. In 
recent years, the study of passive scalars has been associated with 
substantial advances in our theoretical understanding of turbulent flows 
\cite{Kraichnan,kraichnanychen,Falkovich}. Passive scalars share similarities
with three dimensional hydrodynamic turbulence, developing a direct cascade 
(a transfer of variance towards smaller scales with constant flux), and 
intermittency (the spontaneous development of strong gradients at small 
scales). As in the study of Navier-Stokes turbulence, some topics of interest 
include the persistence of anisotropy at small scales 
\cite{sreenivasan,warhaft}, universality, and deviations from scale-invariance 
associated with intermittent events. For the passive scalar, significant 
advances to understand these topics have been made in the framework of 
the Kraichnan model \cite{Kraichnan1,Kraichnan,falkovich96}. For 
a random delta-correlated in time velocity field, the scaling exponents of 
the passive scalar were obtained and shown to have anomalous (intermittent) 
behavior but to be universal. Numerical simulations and experiments showed 
good agreement with these results \cite{kraichnanychen}. The results were 
later also extended to consider the behavior of probability density 
functions, or differences between passive and active scalars (see, e.g., 
\cite{Cencini}).

Passive scalars in rotating turbulence have received less attention. It is 
well known that one of the effects of rotation is to modify the energy 
transfer, transferring the energy preferentially towards modes perpendicular 
to the rotation axis \cite{Cambon89,waleffe92,cambon97}. This results in a 
two-dimensionalization of the flow, and in the development of an inverse 
energy cascade. High resolution simulations agree with these results, and 
also indicate that intermittency in the velocity field is substantially 
reduced by rotation \cite{Muller07,mininni09,mininnipart1,mininnipart2}. 
Passive scalar transport has been studied in numerical simulations showing 
that its transfer is affected by rotation \cite{yeung,brethouwer}, and in 
experiments \cite{moisy} showing that passive scalars in rotating flows 
still develop anomalous scaling in their structure functions.

In this paper we analyze data from direct numerical simulations of the 
Navier-Stokes equation in a rotating frame, together with the 
advection-diffusion equation for the passive scalar. Spatial resolution is 
$512^{3}$ grid points in a regular periodic grid. Passive scalar is injected 
into an initially homogeneous and isotropic turbulent flow, sustained by an 
external (random) mechanical force. After reaching a steady state, rotation 
is turned on. Two different scales are considered for the injection of 
mechanical energy and passive scalar. In one case, both are injected at the 
largest available scale in the domain. In the other, both are injected at an 
intermediate scale, to allow the mechanical energy develop some inverse 
transfer and to see whether this transfer affects the intermittency of the 
passive scalar. Two different rotation rates are considered, in both cases 
chosen to study the regime of moderate Rossby numbers. Inertial range 
scaling is studied considering the energy and passive scalar spectra and 
fluxes. Velocity and passive scalar structure functions are computed using 
an axisymmetric decomposition, and the corresponding scaling exponents are 
considered to characterize intermittency in each field. Finally, 
probability density functions of increments of the velocity and the passive 
scalar are studied, together with visualizations of the fields. We find 
that the passive scalar is more anisotropic than the velocity field at 
small scales, and follows a spectral law consistent with 
$\sim k_\perp^{-3/2}$, where $k_\perp$ denotes wave vectors 
perpendicular to the rotation axis. While intermittency in the velocity 
field is strongly decreased in the rotating case, the passive scalar is 
still intermittent and its scaling exponents show good agreement with 
Kraichnan's prediction for passive scalar intermittency in two-dimensions.

\section{Setup and theory}

\subsection{Basic equations, code, and simulations}
For an incompressible fluid with uniform mass density in a rotating frame,
the Navier-Stokes equation for the velocity field ${\bf u}$, and the equation 
for the passive scalar ${\theta}$ are,
\begin{equation}
\partial_t {\bf u} + {\bf u}\cdot \nabla {\bf u} = -2{\bf \Omega} \times 
    {\bf u} - \nabla p + \nu \nabla^2 {\bf u} +{\bf f},
\label{eq:NS}
\end{equation}
\begin{equation}
\nabla \cdot {\bf u} =0, 
\label{eq:incompressible}
\end{equation}
\begin{equation}
\partial_t {\theta} + {\bf u}\cdot \nabla {\theta} =  
    \kappa \nabla^2 {\theta} +{\bf \phi},
\label{eq:theta}
\end{equation}
where $p$ is the pressure divided by the mass density, $\nu$ is the kinematic 
viscosity, and $\kappa$ is the scalar diffusivity. Here, ${\bf f}$ is an 
external force that drives the turbulence, ${\phi}$ is the source of the 
scalar field, and ${\bf \Omega} = \Omega \hat{z}$ is the rotation angular 
velocity.

For the analysis in the following sections we use data stemming from direct 
numerical simulations of the above equations. We solve Eqs.~(\ref{eq:NS}), 
(\ref{eq:incompressible}) and (\ref{eq:theta}) using a parallel pseudospectral 
code in a three dimensional domain of size $2\pi$ with periodic boundary 
conditions \cite{Gomez05a,Gomez05b}. The pressure is obtained by taking the 
divergence of Eq.~(\ref{eq:NS}), using the incompressibility condition 
(\ref{eq:incompressible}), and solving the resulting Poisson equation. The 
equations are evolved in time using a second order Runge-Kutta method. The 
code uses the $2/3$-rule for dealiasing, and as a result the maximum 
wave number is $k_{max} = N/3$, where $N$ is the number of grid points 
in each direction. All simulations presented are well resolved, in the sense 
that the dissipation wave numbers $k_\nu$ and $k_\kappa$ are smaller than the 
maximum wave number $k_{max}$ at all times.  

The practice of numerically solving flows in a rotating frame in periodic boxes dates
back to Ref. \cite{Bardina85}. Basically, strict periodicity in all three spatial
directions replaces the hypothesis of homogeneity. With these boundary conditions,
the centrifugal force (not written in Eq. \ref{eq:NS}) can be written as a gradient
and absorved into the pressure, and is thus automatically taken into account 
when the Poisson equation is solved. The rotating flow is thus considered infinite, 
homogeneous, with constant rotation rate $\Omega$, and locally expanded in 
the rotating frame using a Fourier series (in the same way local instabilities 
are studied in unbounded flows). There are however two differences between an
infinite and a periodic flow. A periodic flow is bounded in the sense that eddies larger
than the size of the box ($2\pi$ in our dimensionless units) cannot develop.
This will be important when we consider separation of scales in the inverse 
energy cascade.  Also, there is a discussion concerning whether decoupling
between two-dimensional modes and fast waves takes place in rotating flows.
While first-order decoupling was shown for periodic flows (which have discrete wave numbers)
\cite{Babin96}, for infinite domains with continuous wave numbers decoupling does not hold 
\cite{Cambon2004}. In practice, both idealizations have limitations, as other effects in bounded flows (e.g., Ekman layers) are not present with these boundary conditions.

The runs are characterized by Reynolds, Peclet, and Rossby numbers. The 
Reynolds, Schmidt, and Peclet numbers are defined as usual as
\begin{equation}
R_e=\frac{UL}{\nu},
\end{equation}
\begin{equation}
S_c=\frac{\nu}{\kappa},
\end{equation}
\begin{equation}
P_e=S_c R_e ,
\end{equation}
where $U$ is the r.m.s.velocity, and $L$ is the forcing scale of the flow defined as 
$2\pi/k_F$, with $k_F$ the forcing wave number (when the forcing is applied in a wide 
band of wave numbers, $k_F$ is taken as the minimum of the wave numbers in the band). 
For the simulations, $U \approx 1$, and all runs have $\nu=\kappa$ (i.e., $S_c=1$ and 
$P_e=R_e$).

To characterize the strength of rotation, we use the Rossby number
\begin{equation}
R_o = \frac{U}{2L\Omega}.
\end{equation}
It is also useful to introduce a micro-Rossby number defined as the ratio of the 
r.m.s.~vorticity $\omega_\textrm{rms}$ (${\bf \omega} = \nabla \times {\bf u}$) to the 
background vorticity, 
\begin{equation}
R_\omega = \frac{\omega_\textrm{rms}}{2\Omega}.
\end{equation}
For rotation to be important, the Rossby number $R_o$ should be smaller than one, but 
the micro-Rossby number $R_\omega$ should be close to one or larger. If $R_\omega$ is
much smaller than one, non-linear interactions are rapidly damped by the scrambling 
effect of Rossby waves, resulting in a strong quenching of turbulence by rotation 
\cite{cambon97}.

Several simulations were done with fixed linear spatial resolution ($N=512$), and same 
kinematic viscosity ($\nu=\kappa=6\times10^{-4}$). Parameters for all runs are given in 
Table \ref{table:runs}. The forcing used for the velocity field as well as for the 
passive scalar is a superposition of Fourier modes with random phases, delta-correlated 
in time, and injected at the same wave number $k_F$ for both fields. One set of runs 
(set A) has forcing applied at $k \in [1,2]$ (therefore $k_F=1$, and the simulations
have the largest possible separation of scales in the direct energy cascade), 
while another set has forcing at $k=k_F=3$ (set B). 
This latter choice for the injection wave number leaves some room in spectral space 
for an inverse cascade of energy to develop, but also reduces the Reynolds number 
by a factor of three, as scale separation between injection and dissipation is smaller.  
This results in narrower inertial ranges for all simulations in set B. However, the scale separation
between the box size and the forcing scale will allow us to compare runs with and without inverse energy transfer (respectively, sets B and A), and to see whether this transfer has some effect in the scaling of the passive scalar. 

The procedure followed in the simulations in both sets was the same. A simulation of 
the Navier-Stokes equation with $\Omega=0$ was done first, until reaching an isotropic 
and homogeneous turbulent steady state (this takes approximately ten turnover times). Then, 
the passive scalar was injected, and the run was continued for other ten turnover times 
until reaching a steady state for the passive scalar (these runs correspond to run A1 of 
set A, and run B1 of set B). Finally, rotation was turned on. Different 
values of $\Omega$ were considered, to have similar Rossby numbers in both sets. These 
runs were started from the last snapshot of the velocity and the passive scalar of run 
A1 or B1. Each of these runs was continued for over twenty turnover times. Rossby numbers 
for each run are listed in Table \ref{table:runs}.

\begin{table}
\caption{\label{table:runs}Parameters used in the simulations.  $k_F$ is
         the forcing wave number, $\Omega$ the rotation, $R_o$ the Rossby number
         $R_e$ the Reynolds number, and $R_\omega$ the micro-Rossby 
         number.}
\begin{ruledtabular}
\begin{tabular}{cccccc}
Run &  $k_F$ & $\Omega$ & $R_o$ &     $R_e$         & $R_\omega$  \\
\hline
A1 & 1  &  0      &   $\infty$ &  1000  &   $\infty$  \\
A2 & 1  &  4      &   $0.04$   &  1000  &   $0.9$     \\
B1 & 3  &  0      &   $\infty$ &   240  &   $\infty$  \\
B2 & 3  &  12     &   $0.04$   &   240  &   $0.4$     \\
\end{tabular}
\end{ruledtabular}
\end{table}

\subsection{\label{sect:Analysis}Analysis}

Characterization of the flow and passive scalar anisotropy, scaling laws, and intermittency 
is done considering power spectra, fluxes, structure functions, and probability density 
functions of field increments. 

Isotropic energy spectrum and power spectrum of the scalar field are defined as usual 
(summing the power of all modes in Fourier space over spherical shells), and denoted by 
$E(k)$ and $V(k)$ respectively. Since rotation introduces a preferred direction, 
anisotropies will be characterized in spectral space using the so-called reduced 
spectra. The reduced perpendicular energy spectrum $E(k_{\perp})$ and scalar power spectrum 
$V(k_{\perp})$ result from summing the power of all modes in cylindrical shells of radius 
$k_{\perp}$, with their axis aligned with the rotation axis $\hat{z}$. The reduced 
parallel spectra $E(k_{\parallel})$ and $V(k_{\parallel})$ result from summing the power of 
all modes in planes with $k_z = k_{\parallel}$. Detailed definitions can be found in 
\cite{mininni09}.

Two-dimensional axisymmetric spectra can be also defined (see e.g., \cite{cambon97}), and 
give more detailed information of spectral anisotropy. Instead, we will consider here 
an axisymmetric decomposition of structure functions, which will also give us information 
of intermittency in the velocity and scalar fields. Longitudinal increments of the 
velocity and passive scalar fields are defined as:
\begin{equation}
\delta u({\bf x},{\bf l})=\left[{\bf u}({\bf x}+{\bf l})-{\bf u}({\bf x}) \right] \cdot 
    \frac{{\bf l}}{|{\bf l}|}, 
\label{eq:deltav}
\end{equation}
\begin{equation}
\delta \theta({\bf x},{\bf l})=\left[\theta ({\bf x}+{\bf l})-\theta({\bf x}) \right] 
    \cdot \frac{{\bf l}}{|{\bf l}|}, 
\label{eq:deltat}
\end{equation}
where the increment ${\bf l}$ can point in any direction. Structure functions of order $p$ 
are then defined as
\begin{equation}
S_{p}({\bf l})=\left\langle \delta u^{p}({\bf x},{\bf l})\right\rangle , 
\label{eq:S}
\end{equation}
for the velocity field, and as
\begin{equation}
T_{p}({\bf l})=\left\langle \delta \theta^{p}({\bf x},{\bf l})\right\rangle ,
\label{eq:Sp}
\end{equation}
for the passive scalar field. Here, brackets denote spacial average over all values 
of ${\bf x}$.

These structure functions depend on the direction of the increment. In simulations 
without rotation, the field is isotropic and the $SO(3)$ decomposition was used to 
calculate the isotropic component of these structure functions 
\cite{Arad,biferale2,biferale,martin}. The decomposition was implemented computing 
(and averaging) the structure functions for 146 different directions that cover 
almost isotropically the sphere, using the method described in \cite{taylor}. In runs 
with rotation, given the preferred direction and the axisymmetry associated with it, 
we will be interested in increments parallel and perpendicular to ${\bf \Omega}$. We 
denote increments in these two directions as $l_{\parallel}$ and $l_{\perp}$ respectively. 
We then follow the procedure explained in \cite{mininnipart2} to perform a 
decomposition of structure functions based on the $SO(2)\times \Re$ symmetry group 
(rotations in the $xy$ plane plus translations in the $z$ direction). Velocity and 
passive scalar structure functions were computed using 26 different directions for the 
increments ${\bf l}$, generated by integer multiples of the vectors $(1,0,0)$, 
$(1,1,0)$, $(2,1,0)$, $(3,1,0)$, $(0,1,0)$, $(-1,1,0)$, $(1,2,0)$, $(-2,1,0)$, 
$(-1,2,0)$, $(1,3,0)$, $(-3,1,0)$, $(-1,3,0)$, and $(0,0,1)$ for translations in $z$ 
(all vectors are in units of grid points in the simulations), plus the 13 vectors 
obtained by multiplying them by $-1$. Once these structure functions were calculated, 
the perpendicular structure functions $S_p(l_{\perp})$ and $T_p(l_{\perp})$ 
were obtained by averaging over the $24$ directions in the $xy$ plane, and 
the parallel structure functions $S_p(l_{\parallel})$ and $T_p(l_{\parallel})$ 
were computed directly using the generators in the $z$ direction.

For runs in set A, this procedure was applied to eight snapshots of the velocity and 
passive scalar fields separated by at least one turnover time each, while for runs in 
set B we analyzed six snapshots. For large enough Reynolds number, the structure 
functions are expected to show inertial range scaling, i.e., we expect that for some 
range of scales $S_p\sim l^{\xi_p}$ and  $T_p\sim l^{\zeta_p}$ ($l$ may be replaced 
by $l_\perp$ or $l_\parallel$ in the rotating case), where $\xi_{p}$ and $\zeta_{p}$ 
are, respectively, the scaling exponents of order $p$ of the velocity and scalar 
fields. Scaling exponents shown in the following sections are calculated for all 
the snapshots analyzed in each simulation, and averaged over time. Errors are then 
defined as the mean square error; e.g., for the passive scalar exponents, the error is
\begin{equation}
e_{\zeta_{p}}=\frac{1}{N}\sqrt{\sum_{i=1}^{N}\left(\zeta_{p_{i}}-\overline{\zeta_{p}}\right)^{2}},
\label{eq:error}
\end{equation}
where $N$ is the number of snapshots of the field, $\zeta_{p_{i}}$ is the slope obtained 
from a least square fit for the $i$-th snapshot, and $\overline{\zeta_{p}}$ is the mean
value averaged over all snapshots. The error in the least square calculation of the 
slope for each snapshot is much smaller than this mean square error. Extended 
self-similarity \cite{benzi1,benzi2} is not used to obtain the scaling exponents.

\section{Numerical Results}
\subsection{Spectra and fluxes}

\begin{figure}
\centerline{\includegraphics[width=8.9cm]{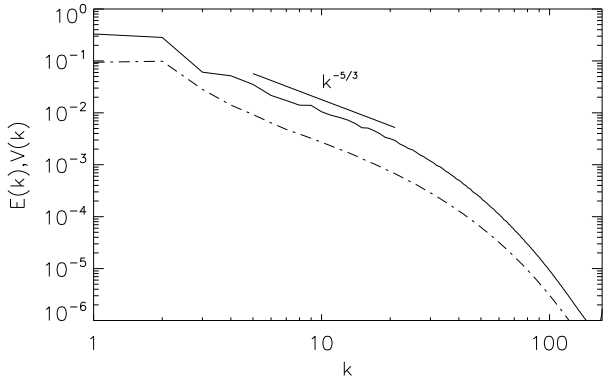}}
\caption{Energy (solid) and passive scalar (dash-dotted) spectrum for run A1. 
Kolmogorov scaling is shown as a reference.}
\label{fig:fig1}
\end{figure}

In the absence of rotation, velocity and passive scalar spectra follow a similar 
inertial range scaling, close to $\sim k^{-5/3}$ as expected from Kolmogorov 
phenomenology, except for bottleneck and possible intermittency corrections 
(the latter only visible in the spectrum at higher resolution \cite{Kaneda,Yeung05}).  
As an example, Figure \ref{fig:fig1} shows the energy and passive scalar spectra 
for run A1 in the turbulent steady state. The slope indicates a $\sim k^{-5/3}$ 
scaling law.

\begin{figure}
\centerline{\includegraphics[width=8.9cm]{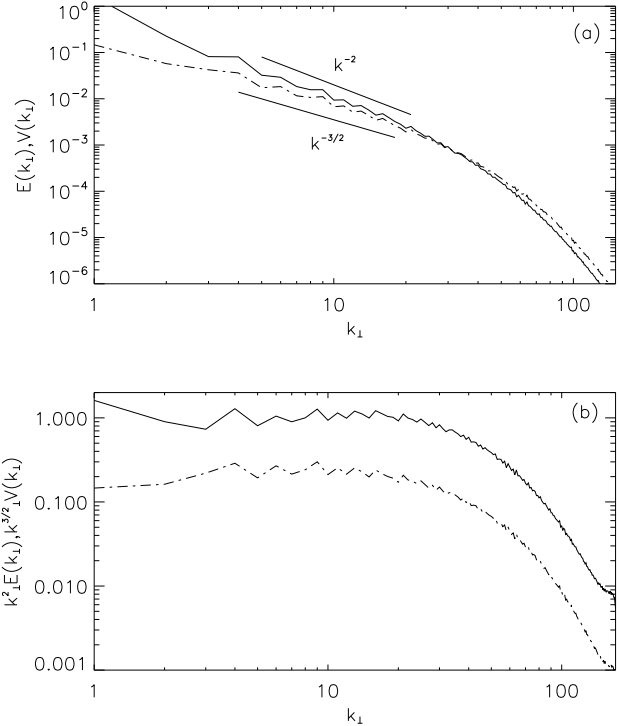}}
\caption{(a) Energy (solid) and passive scalar (dash-dotted) reduced perpendicular spectrum
for run A2. Slopes are shown as a reference. (b) Reduced perpendicular 
energy spectrum compensated by $k_\perp^{-2}$, and passive scalar power spectrum 
compensated by $k_\perp^{-3/2}$ for the same run.}
\label{fig:fig2}
\end{figure}

\begin{figure}
\centerline{\includegraphics[width=8.9cm]{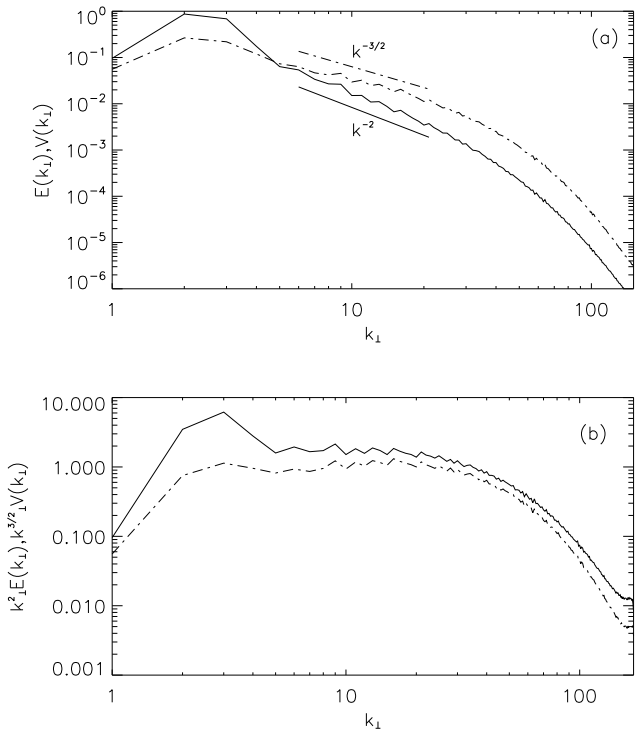}}
\caption{(a) Reduced perpendicular energy (solid) and passive scalar (dash-dotted) 
spectra for run B2. (b) Reduced perpendicular energy spectrum compensated by 
$k_\perp^{-2}$ and reduced perpendicular passive scalar power spectrum compensated by 
$k_\perp^{-3/2}$.}
\label{fig:fig3}
\end{figure}

Once rotation is turned on, the spectral scaling of the passive scalar and velocity changes. 
Figure \ref{fig:fig2}(a) shows the reduced perpendicular energy spectrum 
$E(k_{\perp})$ for run A2, together with the reduced perpendicular passive scalar power 
spectrum $V(k_{\perp})$.  The spectral scaling of the velocity field changes 
and gets closer to the expected $E(k_\perp) \sim k_\perp^{-2}$ \cite{Zhou95,Muller07},
while the passive scalar is close to $\sim k_{\perp}^{-3/2}$ scaling. 
This can be further confirmed in the compensated spectra shown in \ref{fig:fig2}(b). 
Figure \ref{fig:fig3} shows the reduced perpendicular energy and passive scalar spectra for 
run B2; $E(k_{\perp})$ is also close to $\sim k_{\perp}^{-2}$, and $V(k_{\perp})$ 
again is close to $k_{\perp}^{-3/2}$. Note run B2 is forced at $k_F=3$, 
and as already mentioned, in the presence of rotation there is some room in spectral space for 
inverse transfer of energy towards larger scales (indeed, at much later times the energy 
spectrum in this run peaks at $k=1$). However, the scaling of the passive scalar 
seems to be insensitive to the development of such an inverse energy transfer. Since both 
sets of runs show similar scaling, in the following we will concentrate on runs in set 
A (which have a larger scale separation between injection and dissipation, and 
therefore a better resolved direct cascade range), and only resort to runs in set B 
for comparisons.

\begin{figure}
\centerline{\includegraphics[width=8.9cm]{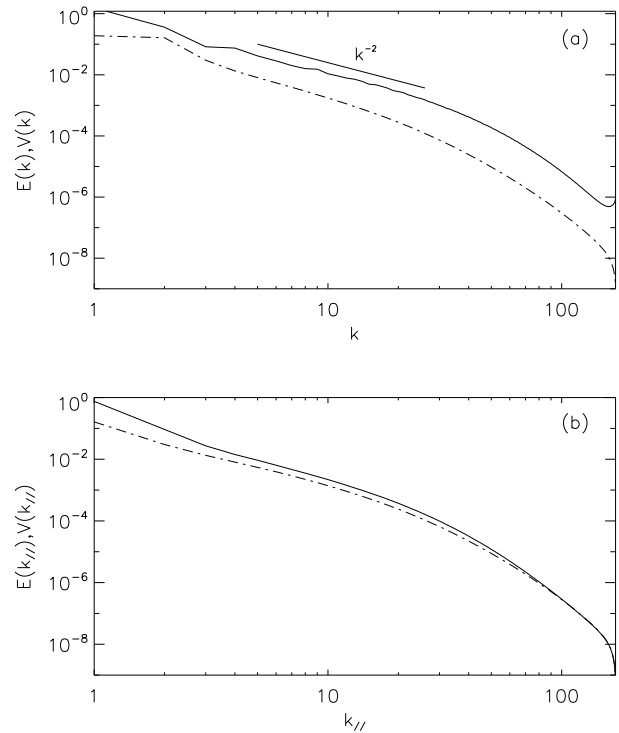}}
\caption{(a) Isotropic energy (solid) and passive scalar (dash-dotted) spectra for run A2. 
(b) Reduced parallel energy (solid) and passive scalar 
(dash-dotted) spectra for the same run.}
\label{fig:fig4}

\end{figure}

Unlike the scaling laws observed in the reduced perpendicular spectra of runs with 
rotation, isotropic and reduced parallel spectra show less clear behavior. Figure 
\ref{fig:fig4}(a) shows the isotropic energy and passive scalar spectra for run A2. 
The isotropic energy spectrum shows a scaling law similar to $k^{-2}$. However, the 
isotropic passive scalar power spectrum shows a steeper spectrum with no clear inertial 
range, hinting at a more anisotropic spectral distribution than the energy at scales smaller 
than the forcing scale. In the case of the reduced parallel spectra, no clear power law 
can be identified for $E(k_{\parallel})$ and $V(k_{\parallel})$; see 
Fig.~\ref{fig:fig4}(b). 

Several quantities can be used to measure the development of anisotropies
in a rotating flow \cite{Cambon89,bartello,cambon97}. As an example, the ratio 
of energy in all modes with $k_{\parallel}=0$ to the total energy, i.e., 
$E(k_\parallel =0)/E$, can be used to characterize large scale anisotropy \cite{mininni09}. 
For a purely two-dimensional flow, this ratio is equal to one. For the passive scalar, 
the equivalent quantity $V(k_\parallel =0)/V$ can also be used. In the simulations without 
rotation, the mean values of $E(k_\parallel =0)/E$ and $V(k_\parallel =0)/V$ are close to 
$0.1$. As can be seen in Table \ref{table:aniso}, these ratios increase when rotation 
is turned on, with $E(k_\parallel =0)/E$ larger than $V(k_\parallel =0)/V$ in all cases. 
This suggests that, at large scales, the velocity field is more anisotropic than the 
passive scalar field.

To quantify small-scale anisotropy, the Shebalin angles can be used instead. These were 
originally introduced to study anisotropy in magnetohydrodynamic turbulence 
\cite{shebalin,matthaeus}, and for the velocity are given by
\begin{equation}
\tan^{2}(\alpha_u)= 2\lim_{l\to0}\frac{S_2(l_\perp)}{S_2(l_\parallel)}=
  2\dfrac{\sum_{k_\perp}k_{\perp}^2 E(k_\perp)}{\sum_{k_\parallel}k_{\parallel}^2 E(k_\parallel)}.
\end{equation}
The angle $\alpha_{u}$ gives a global measure of small-scale anisotropy, with a value 
of $\tan^{2}(\alpha_{u})=2$ corresponding to an isotropic flow. The square tangent of the 
Shebalin angles for the velocity and passive scalar fields, respectively, 
$\tan^{2}(\alpha_u)$ and  $\tan^{2}(\alpha_\theta)$, are given in Table \ref{table:aniso} 
for runs with rotation.
The values of $\tan^{2}(\alpha_\theta)$ are larger than the values of $\tan^{2}(\alpha_u)$,
indicating that the passive scalar field is more anisotropic than the velocity field 
in the small scales. This slower recovery of small-scale isotropy by the passive 
scalar, when compared with the velocity field, has already been reported in the 
case of non-rotating turbulence, see e.g., \cite{warhaft}, but not in the rotating 
case to the best of our knowledge.

\begin{table}
\caption{\label{table:aniso}Anisotropy in runs with rotation.
    $E(k_\parallel =0)/E$ is the ratio of energy in all modes with $k_{\parallel}=0$
    to the total energy, $V(k_\parallel =0)/V$ is the ratio of scalar variance in 
    all modes with $k_{\parallel}=0$ to the total scalar variance, 
    $\tan^{2}(\alpha_u)$ is the square tangent of the Shebalin angle for the 
    velocity field, and $\tan^{2}(\alpha_\theta)$ is the square tangent of the 
    Shebalin angle for the passive scalar.}
\begin{ruledtabular}
\begin{tabular}{ccccc}
Run & $E(k_\parallel =0)/E$ & $V(k_\parallel =0)/V$ & $\tan^{2}(\alpha_u)$ & 
$\tan^{2}(\alpha_\theta)$   \\
\hline
A2 &  0.5      &   0.4   &  13   & 20    \\
B2 &  0.2      &   0.1   &  14   & 50    \\
\end{tabular}
\end{ruledtabular}
\end{table}

\begin{figure}
\centerline{\includegraphics[width=8.9cm]{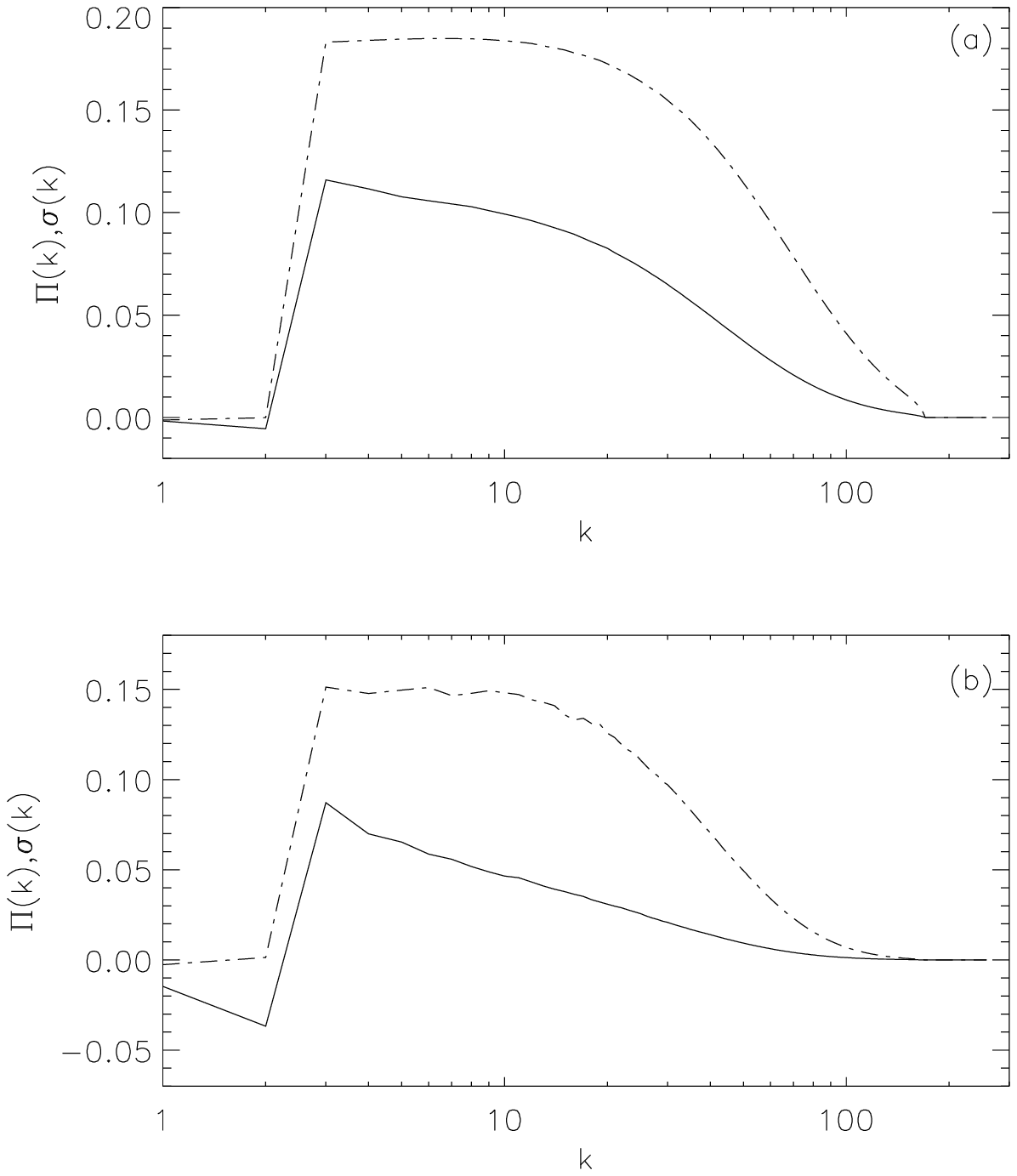}}
\caption{(a) Energy flux $\Pi(k)$ (solid) and passive scalar flux $\sigma(k)$ 
(dash-dotted) for run B1. (b) Same for run B2.}
\label{fig:fig5}
\end{figure}

The growth of energy at $k=1$, and the flux of energy towards large scales, 
associated with the two-dimensionalization of the flow resulting from rotation, 
can only be observed in run B2 (see Fig.~\ref{fig:fig5}). As mentioned before, 
this run has some room in spectral space between the largest wave number in 
the box and the forcing wave number. Figure \ref{fig:fig5} shows the energy and 
passive scalar fluxes in runs B1 and B2 (without and with rotation, respectively) 
as a comparison. In the absence of rotation, the energy flux shows a direct 
cascade range (a range of positive and approximately constant flux), while 
the energy flux is negligible for wave numbers smaller than the forcing wave number 
($k \le k_{F}$). The passive scalar flux shows a similar behavior. In the presence 
of rotation, the energy flux becomes negative for $k \le k_{F}$, indicating energy 
is transferred towards scales larger than the forcing scale, and the flux towards 
smaller scales decreases. Scale separation is too small to talk of an inverse cascade of energy 
with constant flux, but computational limitations in our attempt to have a well 
resolved direct cascade of passive scalar limits our ability to resolve an inverse 
energy cascade. For the passive scalar, no flux toward larger scales is 
observed, and the amplitude of the positive (direct) flux decreases by a small 
fraction.

Based on these results, we can put forward a phenomenological argument considering 
the effect of rotation, to obtain the inertial range spectrum of passive scalar 
observed in the simulations. From  Eq.~(\ref{eq:theta}), we 
can estimate the flux of passive scalar as
\begin{equation}
\sigma \sim \frac{u_{l_\perp} \theta_{l_\perp}^{2}}{l_\perp},
\label{eq:sigma}
\end{equation}
where $u_{l_\perp}$ is the characteristic velocity at the scale $l_\perp$, and 
$\theta_{l_\perp}$ is the characteristic concentration of passive scalar at the same 
scale. Using that the passive scalar flux is constant in the inertial range, then 
$\theta_{l_\perp}$ scales as $\theta_{l_\perp}^{2}\sim \sigma l_\perp /u_{l_\perp}$. If 
the energy spectrum in the direct energy cascade range follows a power law 
$E(k_\perp) \sim k_{\perp}^{-2}$, then in the absence of intermittency corrections 
$u_{l_\perp} \sim l_\perp^{1/2}$, and
\begin{equation}
V(k_{\perp}) \sim \sigma k_{\perp}^{-3/2} .
\label{eq:Vk}
\end{equation}


\subsection{Structure functions and scaling exponents}

We now consider isotropic and axisymmetric velocity and passive scalar structure 
functions for the simulations described above, using the method explained in 
Sec.~\ref{sect:Analysis}. 

\begin{figure}
\centerline{\includegraphics[width=8.9cm]{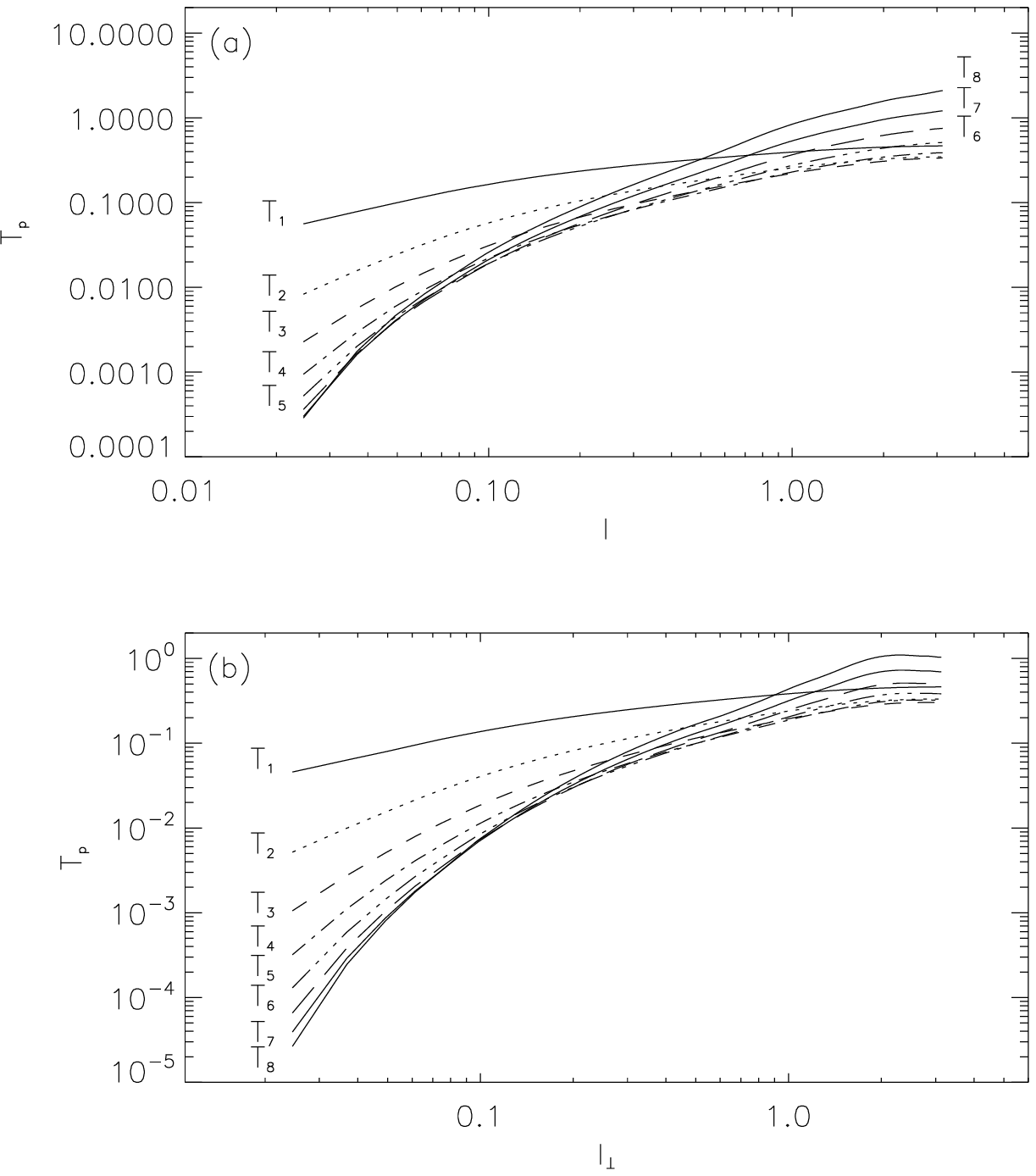}}
\caption{(a) Isotropic structure functions for the passive scalar up to 8th order 
in run A1, at $t=7$. (b) Axisymmetric (only for $l_\perp$ increments) structure 
functions for the passive scalar up to 8th order in run A2, at $t=18$.}
\label{fig:fig6}
\end{figure}

Figure \ref{fig:fig6}(a) shows the result of computing the passive scalar structure 
functions in run A1 (isotropic) and A2 (axisymmetric, only structure functions with 
$l_\perp$ increments are shown), for one instantaneous snapshot of the scalar field 
(at $t=7$ in run A1, and at $t=18$ in run A2). The structure functions show a range 
of scales with approximately power law scaling at intermediate scales, and at the 
smallest scales approach the $\sim l^p$ scaling expected for a smooth field in the 
dissipative range.

\begin{figure}
\centerline{\includegraphics[width=8.9cm]{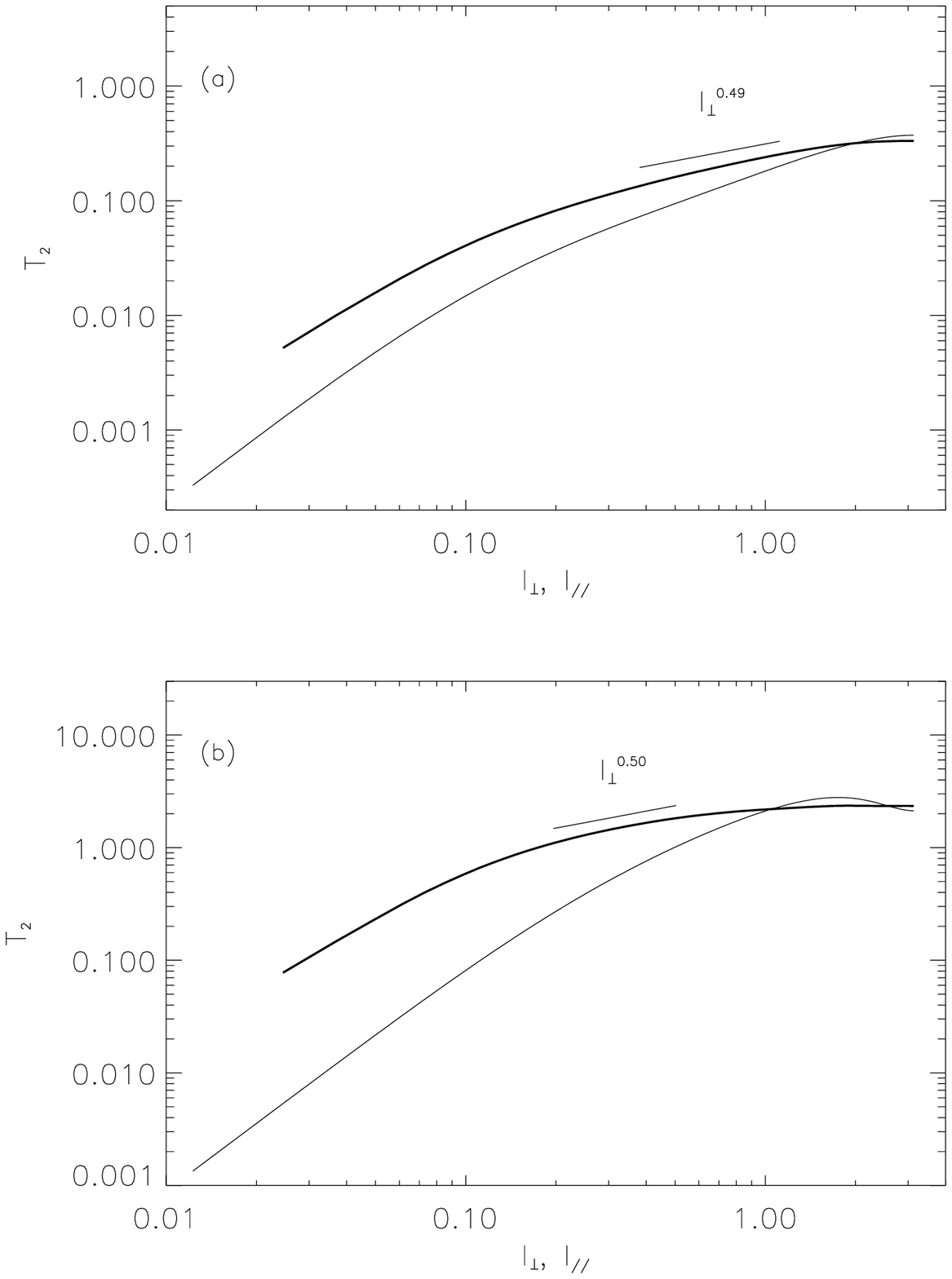}}
\caption{Axisymmetric second order structure functions for the passive scalar: 
(a) run A2 with $\Omega=4$ at $t=18$, and 
(b) run B2 with $\Omega=12$ at $t=14$. 
$T_2(l_\perp)$ corresponds to dashed lines and $T_2(l_\parallel)$ to full lines. 
Slopes indicated as references correspond to the time average of the scaling exponents, 
obtained from a best fit in the inertial range of the structure functions at different 
times.}
\label{fig:fig7}
\end{figure}

Figure \ref{fig:fig7} shows a detail for rotating runs A2 and B2 of the passive 
scalar second order axisymmetric structure functions in the perpendicular direction 
$T_2(l_\perp)$ (i.e., averaged in all directions in the $xy$ plane), and in the 
parallel direction, $T_2(l_\parallel)$. Stronger anisotropy is observed 
at small scales, manifested as a different 
(steeper) dependence of $T_2(l_\parallel)$ with the spatial increment when compared 
with $T_2(l_\perp)$, and as a much smaller amplitude of $T_2(l_\parallel)$ at small 
scales. 

An inertial range with power law-scaling can be 
identified at intermediate scales in $T_{2}(l_{\perp})$, but not so clearly in 
$T_{2}(l_{\parallel})$ (this is specially true for run B2, where the parallel 
structure function shows only the smooth scaling $~l_{\parallel}^2$). This is 
consistent with results shown in the previous section for the reduced passive 
scalar spectra. Inertial range with power-law scaling was observed for the 
reduced perpendicular spectrum $V(k_{\perp})$, while no clear scaling could be 
identified for $V(k_{\parallel})$. Note also that at scales close to the forcing 
scale or larger, the passive scalar distribution always seems to be more isotropic, 
in agreement with the results shown in Table \ref{table:aniso}.

The slopes indicated as a reference in Fig.~\ref{fig:fig7} correspond to the time 
average of the second order scaling exponent, obtained from a best fit in the inertial 
range of all structure functions at different times available for each run (eight 
snapshots for run A2, and six snapshots for run B2). The second order scaling 
exponents (in the perpendicular direction) for the passive scalar thus obtained are 
$\zeta_{2}\approx0.49\pm0.01$ in run A2, 
and $\zeta_{2}\approx0.50\pm0.01$ in run B2. 

The values obtained for $\zeta_{2}$ are in good agreement with the 
$V(k_{\perp})\sim k_\perp^{-3/2}$ power law found for the reduced perpendicular 
passive scalar spectrum, which leads on dimensional grounds to 
$T_{2}(l_{\perp})\sim l_{\perp}^{-1/2}$. 

\begin{figure}
\centerline{\includegraphics[width=8.9cm]{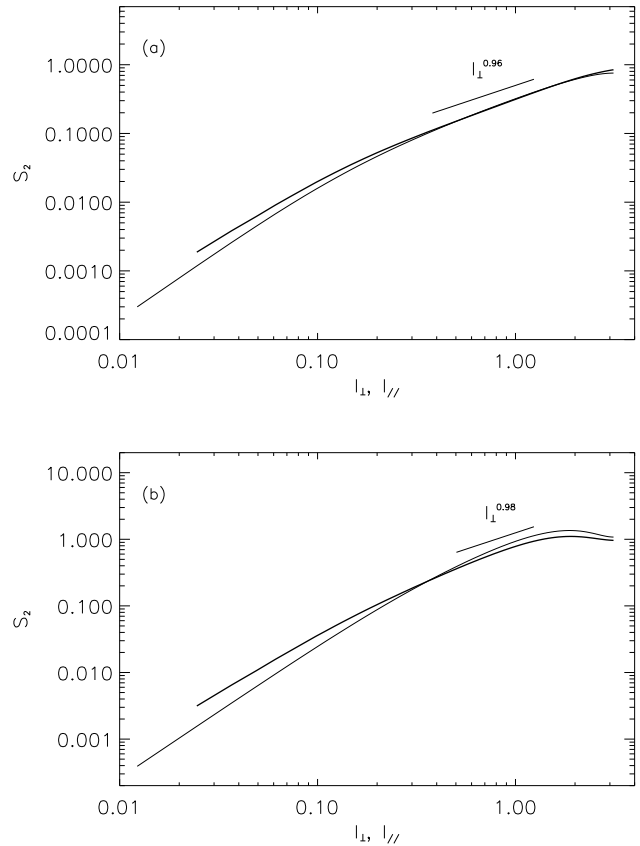}}
\caption{Axisymmetric second order structure functions for the velocity field:
(a) run A2 with $\Omega=4$ at $t=18$, and 
(b) run B2 with $\Omega=12$ at $t=14$.
$S_2(l_\perp)$ corresponds to dashed lines and $S_2(l_\parallel)$ to full lines. 
Slopes indicated as references correspond to the time average of the scaling exponents, 
obtained from a best fit in the inertial range of all structure functions at different 
times.}
\label{fig:fig8}
\end{figure}

Figure \ref{fig:fig8} shows the axisymmetric second order structure functions for 
the velocity field, for runs with rotation at late times. 
Small-scale anisotropy develops, and is more clearly seen in run B2 with $\Omega=12$. 
At large scales, stronger 
correlations are observed in the parallel direction in this latter case, which is to 
be expected as this simulation has some separation between the largest scale 
available in the box and the forcing scale, and the flow becomes quasi-2D at large 
scales. Comparing these structure functions with the ones for the passive scalar in 
Fig.~\ref{fig:fig7}, it becomes apparent that at the Rossby number considered, the 
small-scale anisotropy associated with rotation is stronger for the passive scalar 
than for the velocity field.

Inertial range with power law scaling is also observed for velocity structure 
functions. For runs A2 and B2, 
time averaged second order scaling exponents, corresponding to 
scaling $S_{2}(l_{\perp})\sim l_{\perp}^{\xi_{2}}$, are indicated as a reference by 
the slopes in Figs.~\ref{fig:fig8}(a) and ~\ref{fig:fig8}(b). Second order scaling exponents are 
$\xi_{2}\approx0.96\pm0.01$ for run A2, 
and $\xi_{2}\approx0.98\pm0.01$ for run B2. These values are consistent with the 
inertial range scaling found for the perpendicular energy spectrum in these runs, 
$E(k_{\perp})\sim k_{\perp}^{-2}$, which leads to 
$S_{2}(l_{\perp})\sim l_{\perp}^{-1}$.

From the curves in Fig.~\ref{fig:fig6}, scaling exponents can be computed for higher 
orders. Velocity and passive scalar exponents in the direct cascade range were computed 
for runs in set A up to the seventh order. For runs in set B, scaling exponents 
were only computed up to the sixth order, as those runs have smaller scale separation 
between forcing and dissipation.

\begin{figure}
\centerline{\includegraphics[width=8.9cm]{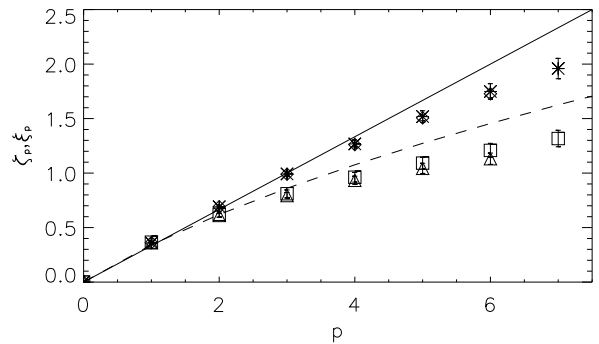}}
\caption{Scaling exponents (with error bars) as a function of the order $p$, for 
the velocity (stars for run A1 and diamonds for run B1), and the passive scalar 
(squares for run A1 and triangles for run B1). The solid line corresponds to 
Kolmogorov scaling $\xi_{p}=p/3$, and the dashed line corresponds to Kraichnan's 
linear ansatz with $d=3$ and $\zeta_2 =0.62$.}
\label{fig:fig9}
\end{figure}

Figure \ref{fig:fig9} shows the velocity and scalar exponents for runs A1 and B1 (no 
rotation). For the velocity field, exponents in both runs are similar. 
The third order velocity field exponent is 
$\xi_{3}=0.99\pm0.02$ for run A1, and $\xi_3=0.98\pm0.01$ for 
run B1, in good agreement with the value of $1$ expected from the $4/5$-law for 
isotropic and homogeneous turbulence. The velocity 
field exponents display the well known deviations from the linear ($\xi_p = p/3$) 
Kolmogorov scaling associated with intermittency. This deviation from strict scale 
invariance is often quantified in terms of the intermittency exponent 
$\mu=2\xi_{3}-\xi_{6}$, which for these runs is $\mu=0.24\pm0.12$ for run A1, and 
$\mu=0.25\pm0.06$ for run B1. The higher order velocity exponents obtained 
from these runs are also consistent with previous results for non-rotating turbulence 
at large Reynolds numbers \cite{Antonia,Leveque,martin}.

The passive scalar exponents for these two runs also display similar values, which 
are consistent (within error bars) with the values of the velocity field exponents 
for orders $p=1$ and 2, and show larger deviations from the Kolmogorov scale invariant 
prediction for larger values of $p$. This is a well known result which indicates that 
the passive scalar in isotropic and homogeneous turbulence is more intermittent than 
the velocity field \cite{Kraichnan,falkovich96}. For advection and diffusion of a 
passive scalar in a random, delta-correlated in time velocity field in a space with 
dimensionality $d$, Kraichnan obtained an expression for the passive scalar scaling 
exponents of all orders as a function of the second order exponent $\zeta_2$
\cite{Kraichnan},
\begin{equation}
\zeta_p = \frac{1}{2}\left[ \sqrt{2d\zeta_2 p+(d-\zeta_2)^{2}}+(d-\zeta_2)\right].
\label{eq:ansatz}
\end{equation}
This prediction is also shown in Fig.~\ref{fig:fig9}, with the exponents computed 
using the value of $\zeta_2$ obtained from the simulations, and for $d=3$. 
Deviations from this model, which start at $p=4$, are small when compared with the 
deviations of the data from the linear scaling. Similar deviations were also already 
reported in previous numerical simulations of passive scalar transport in turbulent 
flows \cite{kraichnanychen}.

\begin{figure}
\centerline{\includegraphics[width=8.9cm]{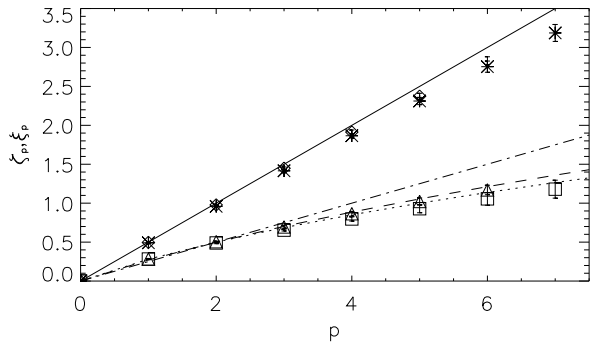}}
\caption{Scaling exponents (with error bars) as a function of the order $p$, for 
the velocity (stars for run A2 and diamonds for run B2) and passive scalar (squares 
for run A2 and triangles for run B2). The solid line corresponds to linear scaling 
$\xi_{p}=p/2$, the dash-dotted line corresponds to linear scaling $\zeta_{p}=p/4$, 
and the dotted and dashed lines correspond respectively to the linear ansatz 
with $d=2$ and $d=3$ using $\zeta_{2}=0.5$.}
\label{fig:fig10}
\end{figure}

In the presence of rotation and for small Rossby number, velocity scaling 
exponents of a strictly self-similar flow are expected to follow a linear law 
$\xi_p=p/2$, if the energy spectrum scales as $\sim k_\perp^{-2}$. 
For runs A2 and B2, velocity scaling exponents are closer to the $p/2$ linear scaling, 
and display significantly smaller deviations from the straight line for the higher 
order exponents than in the non-rotating case (see Fig.~\ref{fig:fig10}). Velocity 
scaling exponents are similar for both runs with rotation. The intermittency 
coefficient in runs A2 and B2 is $\mu=0.1\pm0.1$. As reported before 
\cite{Baraud,Muller07,Seiwert,mininni09,mininnipart2}, for large enough rotation 
intermittency is substantially decreased.

Figure \ref{fig:fig10} also shows the scaling exponents 
for the passive scalar in runs with rotation.
In the scale invariant (non-intermittent) case, a $\zeta_{p}=p/4$ 
linear law is expected for passive scalar exponents, if the 
second order exponent is $\zeta_2=1/2$. Deviations from this linear scaling are 
clearly visible in all runs (see, e.g, Fig.~\ref{fig:fig10}). For runs A2 and B2, 
Kraichnan's linear ansatz (\ref{eq:ansatz}) adjusts very accurately the numerical 
data with $\zeta_2=0.5$ and $d=2$. The good agreement with the 
model with $d=2$ for runs with small Rossby number 
again points to a strong bi-dimensionalization of the passive scalar 
distribution in the presence of rotation.

Intermittency of the passive scalar  decreases with the Rossby number, 
although less than for velocity. The intermittency coefficient for the passive 
scalar, $\mu_{s}=2\zeta_{3}-\zeta_{6}$, is $\mu_{s}=0.4\pm0.1$ for runs
A1 and B1 ($\Omega=0$), and $\mu_{s}=0.2\pm0.1$ 
and $\mu_{s}=0.22\pm0.09$ for runs A2 and B2 respectively ($\Omega \neq 0$).

\subsection{Probability density functions}

\begin{figure}
\centerline{\includegraphics[width=8.9cm]{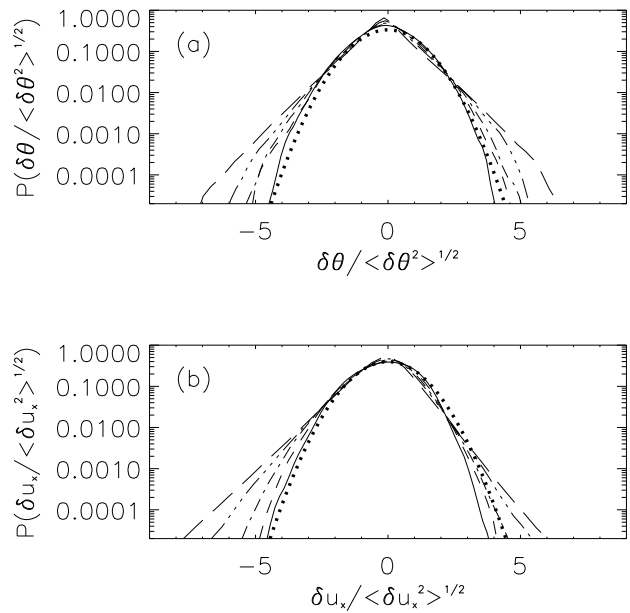}}
\caption{Probability density functions for passive scalar and velocity 
increments in run A1, for five different spatial increments $l=1.6$ (solid), 
$0.8$ (dashed), $0.4$ (dashed-dotted), $0.2$ (dashed-dotted-dotted), 
and $0.1$ (long dashes). (a) Passive scalar, and (b) $x$-component of the 
velocity field. In both cases the dotted curve represents a Gaussian distribution 
with unit variance. Smaller increments have stronger non-Gaussian tails.}
\label{fig:fig11}
\end{figure}
\begin{figure}
\centerline{\includegraphics[width=8.9cm]{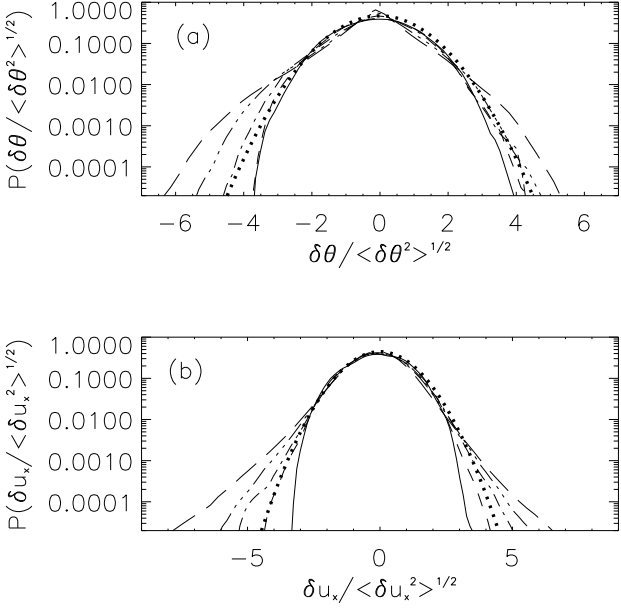}}
\caption{Probability density functions for passive scalar and velocity 
increments in run A2, for five different spatial increments $l=1.6$ (solid), 
$0.8$ (dashed), $0.4$ (dashed-dotted), $0.2$ (dashed-dotted-dotted), 
and $0.1$ (long dashes). 
(a) Passive scalar, and (b) $x$-component of the velocity field.
A Gaussian curve with unit variance is indicated by the dotted curve.
Again, as intervals are decreased, curves depart more and more from the 
Gaussian distribution.}
\label{fig:fig12}
\end{figure}

In this section we consider probability density functions (PDFs) for longitudinal 
increments and derivatives of the $x$-component of the velocity field, as 
well as for the passive scalar. In all cases, curves shown are normalized by their 
variance, and a Gaussian curve with unit variance is shown as a reference. In a 
scale invariant flow the PDFs of the velocity and scalar increments
are expected to be Gaussian and to 
collapse to a single curve when properly normalized. On the other hand, in an 
intermittent flow, probability density functions are expected to have strong 
non-Gaussian tails with larger amplitude as smaller spatial increments are considered. 
These tails are associated with the larger-than-Gaussian probability 
of extreme events to occur, which is the signature of intermittency.

Figure \ref{fig:fig11} shows the PDFs of the velocity and the passive scalar increments 
for four different values of the spatial increment $l=1.6$, $0.8$, $0.4$, $0.2$, and 
$0.1$, for run A1 ($\Omega=0$). As a reference, the forcing scale in this runs is 
$\approx 2\pi$ and the dissipative scale is $\approx0.05$; $l=0.8$ and $0.4$ clearly
correspond to scales in the inertial range. The PDFs of velocity 
and passive scalar increments for $l=1.6$ are close to Gaussian, while for smaller 
spatial increments non-Gaussian tails develop.  

\begin{figure}
\centerline{\includegraphics[width=8.9cm]{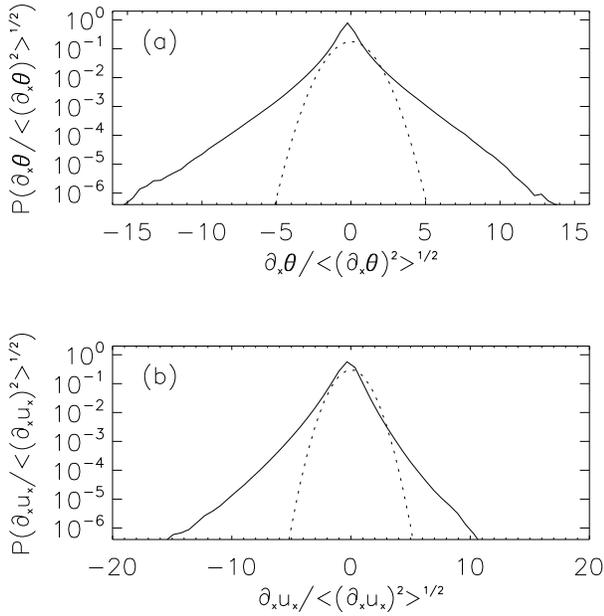}}
\caption{Probability density functions for velocity and passive scalar derivatives
in the $x$ direction for run A1. (a) Passive scalar, and 
(b) $x$-component of the velocity field. Dotted curves represent a Gaussian 
distribution with unit variance.}
\label{fig:fig13}
\end{figure}
\begin{figure}
\centerline{\includegraphics[width=8.9cm]{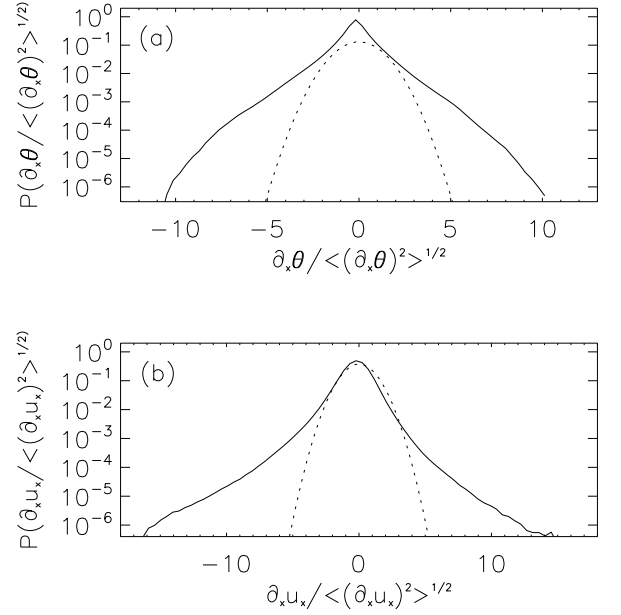}}
\caption{Probability density functions for velocity and passive scalar derivatives
in the $x$ direction for run A2. (a) Passive scalar, and (b) $x$-component of the 
velocity field. Dotted curves represent a Gaussian distribution with unit variance.}
\label{fig:fig14}
\end{figure}

Figure \ref{fig:fig12} shows the PDFs of passive scalar and velocity 
increments for rotating run A2. For this run deviations from Gaussianity in passive 
scalar PDFs start at $l=0.4$, and strong asymmetry is observed for 
$l=0.4, 0.2$ and $0.1$. Velocity increments also deviate from Gaussian 
distribution at $l=0.4$. 
The asymmetry in the tails of the PDFs of passive scalar increments becomes more 
important when rotation is present. This is in good agreement with results found 
in the previous section, which indicate an increment in small scale anisotropy with 
rotation.

\begin{table}
\caption{\label{table:SyK}Skewness ($S$) and kurtosis ($K$) for runs A1 and A2, for 
different quantities (for scalar or field increments, $l$ is the spatial increment 
considered).}
\begin{ruledtabular}
\begin{tabular}{cccccc}
Quantity               &   $l$  & $S$(A1) & $S$(A2) & $K$(A1) & $K$(A2) \\
\hline
$\partial_{x} \theta$  &   --   & $0.04$  & $-0.15$ & $12.11$ & $9.01$  \\
$\delta \theta$        &  $1.6$ & $-0.06$ & $0.01$  & $3.1$   & $2.8$   \\
$\delta \theta$        &  $0.8$ & $0.006$ & $-0.04$ & $3.4$   & $2.9$   \\
$\delta \theta$        &  $0.4$ & $0.005$ & $-0.2$  & $4.0$   & $3.3$   \\
$\delta \theta$        &  $0.2$ & $0.02$  & $-0.2$  & $4.6$   & $4.2$   \\
$\delta \theta$        &  $0.1$ & $0.03$  & $-0.2$  & $6.0$   & $5.4$   \\
\hline
$\partial_{x}u_{x}$    &   --   & $-0.65$ & $-0.44$ & $8.15$  & $7.96$  \\
$\delta u_{x}$         &  $1.6$ & $-0.19$ & $-0.08$ & $2.8$   & $2.6$   \\
$\delta u_{x}$         &  $0.8$ & $-0.25$ & $-0.08$ & $3.2$   & $2.8$   \\
$\delta u_{x}$         &  $0.4$ & $-0.31$ & $-0.02$ & $3.6$   & $3.1$   \\
$\delta u_{x}$         &  $0.2$ & $-0.36$ & $-0.05$ & $4.3$   & $3.6$   \\
$\delta u_{x}$         &  $0.1$ & $-0.44$ & $-0.2$  & $5.3$   & $4.6$   \\
\end{tabular}
\end{ruledtabular}
\end{table}

\begin{figure}
\centerline{\includegraphics[width=6.5cm]{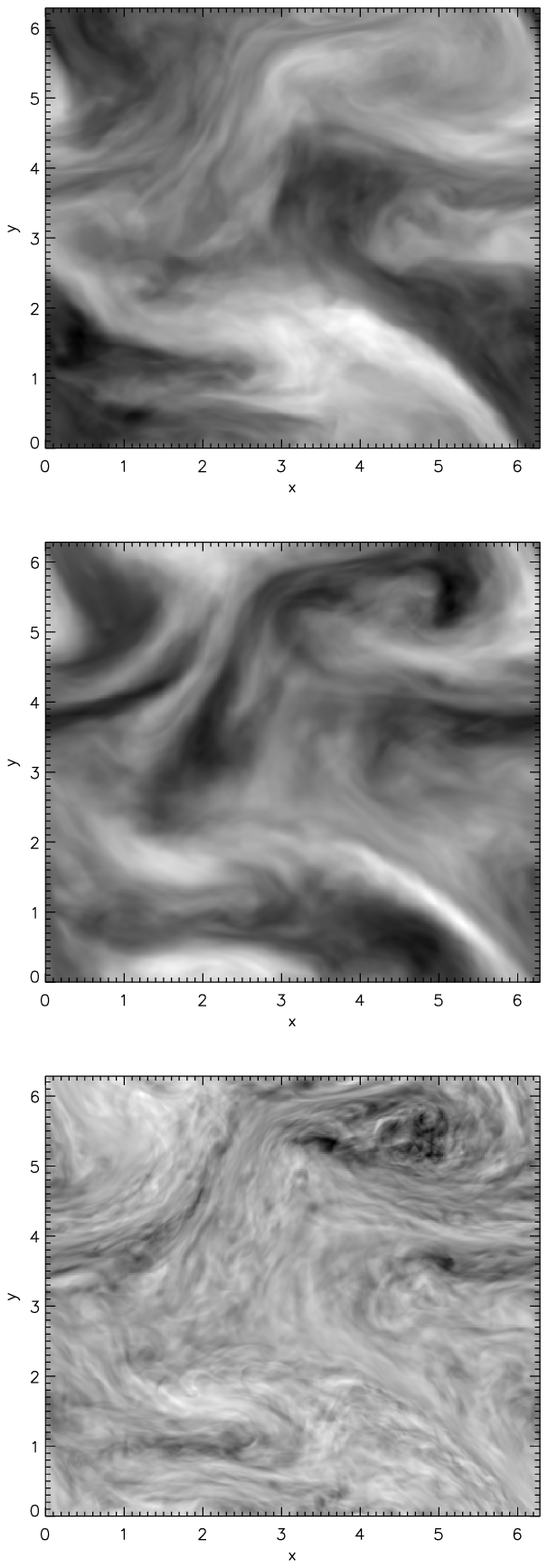}}
\caption{Horizontal average of the passive scalar (top), vertical velocity (middle) and 
vertical vorticity (bottom) for one snapshot of the fields at late times in run A2.}
\label{fig:fig15}
\end{figure}

In figure \ref{fig:fig13} we show the PDFs of velocity and passive scalar derivatives 
(respectively $\partial_{x}u_{x}$ and $\partial_{x} \theta$), in  run A1 ($\Omega=0$).
Deviations from Gaussian statistics are observed for both derivatives. 
The PDFs of the derivatives $\partial_{x} \theta$ and $\partial_{x}u_{x}$ for 
run A2 ($\Omega \neq 0$) shown in Fig.~\ref{fig:fig14} also present strong 
deviations from Gaussianity and asymmetry for $\partial_{x} \theta$. Calculation 
of the skewness for the PDFs of $\partial_{x} \theta$ shows a value of 
$S_{\theta}=0.04$ for run A1, and $S_{\theta}=-0.15$ for run A2. For 
velocity gradients, skewness is $S_{u}=-0.65$ for run A1, and $S_{u}=-0.44$ for run 
A2.

Since differences in the tails of the PDFs in the rotating and non-rotating cases 
may not be easily appreciated in the figures, Table \ref{table:SyK} lists the 
skewness and kurtosis for several quantities computed for runs A1 and A2. For 
the scalar field, the skewness is close to zero for all quantities when $\Omega=0$, 
but it becomes negative and with larger absolute values in the presence of rotation. 
The kurtosis of all quantities studied decreases in the rotating case. For the 
velocity field, the skewness is negative in all cases, but substantially decreases 
in the presence of rotation. As for the passive scalar, kurtosis of quantities 
associated with the velocity field also decreases with rotation.

Many of the results shown indicate both the passive scalar and the velocity field 
become anisotropic in the runs with rotations. To illustrate this, Fig.~\ref{fig:fig15} 
shows an horizontal average of the passive scalar, of the $z$-component of the 
velocity field, and of the $z$-component of the vorticity, at late times in run A2. 
Some correlations between the distribution of the three quantities can be identified, 
as the large scale configuration is very similar in the three cases. However, 
vertical vorticity develops structures at much smaller scales, as expected. A 
detailed analysis of spatial correlations between the fields, and of spatial 
(turbulent) diffusion of the passive scalar, is left for future work.

\section{Conclusions}

In this paper we analyzed data from direct numerical simulations of advection 
and diffusion of a passive scalar in a rotating turbulent flow. The results 
were compared with passive scalar dynamics in isotropic and homogeneous 
turbulence. To this end, simulations with spatial resolution of $512^{3}$ grid 
points were performed in a regular periodic grid using a pseudospectral method. 
Turbulence was sustained by an external random mechanical forcing. The 
forcing scale was changed to consider the case where energy was injected at the 
largest available scale in the computational domain, as well as the case where 
a small separation of scales between the forcing and the size of the box 
allowed for some inverse transfer of energy to develop in the rotating case. 
The main focus of our study was in the resulting scaling laws in spectral space, 
and in intermittency as measured from structure functions and scaling exponents.

The most important result of this work is that, in rotating flows at moderate 
Rossby number, all scaling exponents of the passive scalar except for one can 
be determined with good accuracy from Kraichnan's model 
\cite{Kraichnan1,Kraichnan} for a flow with dimensionality $d=2$ (at least up 
to the highest order considered here, and within the error bars of our study). 
The value of $d$ can be understood from the effect of rotation, which transfers 
the energy preferentially towards modes perpendicular to the rotation axis 
\cite{Cambon89,waleffe92,cambon97} and makes the flow quasi-two-dimensional. 
Moreover, all exponents in Kraichnan's model depend on the second order scaling 
exponent for the passive scalar $\zeta_2$. The value of this exponent can be 
determined from simple dimensional arguments, and we showed the value of 
$\zeta_2=0.5$ to be consistent with phenomenology and with the numerical 
results.

The other results include an analysis of scaling of the power spectrum of the 
passive scalar, which in the rotating case scales as $\sim k_\perp^{-3/2}$. 
This result seems to be independent of whether there is scale separation 
between the forcing scale and the largest scale in the box, or not. In the 
former case, it was also observed that while the energy develops an inverse 
transfer in the presence of rotation, the passive scalar does not. Measures of 
anisotropy based on the spectrum and on structure functions also indicate that 
the passive scalar is more anisotropic at small scales (scales smaller than 
the forcing scale) than the velocity, although at larger scales the passive 
scalar is more isotropic than the velocity field.

Finally, probability density functions of derivatives and longitudinal 
increments of the velocity field and of the passive scalar show a decrease 
of intermittency in the rotating case (as indicated, e.g., by a decrease in 
their kurtosis), more pronounced for the velocity than for the passive scalar. 
This is also in agreement with results obtained from the structure functions 
of the scalar and velocity field, as well as from the intermittency exponents 
for each field.

\begin{acknowledgments}
PDM acknowledges support from the Carrera del Investigador Cient\'{\i}fico of 
CONICET. PRI and PDM acknowledge support from grants UBACYT 20020090200692,
PICT 2007-02211, and PIP 11220090100825.
\end{acknowledgments}

\bibliography{ms}

\end{document}